\documentstyle[multicol,prl,aps]{revtex}
\begin{document}
\title{On the Stereochemistry of the Cations\\ in
the Doping Block of Superconducting Copper-Oxides}
\author{J. R\"ohler\cite{byline}}
\address{European Synchrotron Radiation Facility,
B.P. 220, F-38043 Grenoble, France}

\date{November 11, 1995}

\maketitle

\begin{abstract}
Metal-oxygen complexes containing Cu,- Tl-, Hg-, Bi- and Pb-cations are
electronically active in superconducting copper-oxides by stabilizing
single phases with enhanced $T_c$, whereas other metal-oxygen complexes
deteriorate copper-oxide superconductivity. Cu, Tl, Hg, Bi, Pb in their
actual oxidation states are closed shell $d^{10}$ or inert $s^2$ pair
ions. Their electronic configurations have a strong tendency to polarize
the oxygen environment. The closed shell $d$ ions with low lying
$nd^{10}\leftrightarrow nd^9(n+1)s$ excitations form linear complexes
through $d_{z^2}-s$ hybridization polarizing the apical oxygens.
Comparatively low $nd^9(n+1)s$ excitation energies distinguish $\rm
Cu^{1+,3+}, Tl^{3+}, Hg^{2+}$ from other closed shell $d^{10}$ ions
deteriorating copper-oxide superconductivity, {\it e.g.} $\rm Zn^{2+}$.
\end{abstract}

\begin{multicols}{2}

\section{Introduction}
Electronic inhomogenities are an important feature of the
superconducting copper-oxides. Upon doping the $\rm CuO_2$ planes are
separated into hole rich and hole poor regions on a mesoscopic
lengthscale \cite{SigSim}. Phase separation in the doped $\rm CuO_2$
planes renders the microscopic mechanisms of normal electric
conductivity and of superconductivity unconventional. To reduce the
complexity of the problem many approaches to the phase separation
problem neglect the details of the atomic and electronic structure of
the so-called doping block. Its function is thus reduced to an efficient
charge transfer pump.

On the other hand the phenomenology of the atomic structures of the
superconducting copper-oxides shows specific properties of the
doping-blocks beyond their function as charge reservoirs to be relevant
for superconductivity in the $\rm CuO_2$ planes. For example,
inelastic neutron scattering experiments on $\rm RBa_2Cu_3O_{7-\delta}$
($\delta =1$-0) compounds yielded strong evidence for relationships
between phase separation in the planes and ordering of the oxygen
dopants in the chains \cite{MesMut}.

Due to the large variety of doping blocks, various types of
non-stoichiometries, locally distorted clusters, disorder-order
phenomena, variable valence states, and anomalous lattice dynamics, it
seems to be a hopeless undertaking to generalize other electronic
properties than their well established charge transfer capability.
Therefore the observed relationship between phase separation in the
planes and oxygen ordering in the doping block of $\rm
RBa_2Cu_3O_{7-\delta}$ might be interpreted as an exceptional case
arising from the chain structure in this particular compound. However,
recent neutron scattering experiments on superoxygenated $\rm
La_2CuO_{4+\delta}$ show additional evidence for a possible relationship
between phase separation and ordering in the doping block. The dopants
in $\rm La_2CuO_{4+\delta}$ are excess oxygens which occupy an
interstitial site between the La-O layers. The interstitial excess
oxygens turn out to order in intercalating layers \cite{WelYam} instead
to distribute themselves randomly, as assumed so far.

Not at least these recent findings on relationships between phase
separation in the planes and ordering phenomena in the doping block
strengthen the need for a more detailed understanding of the physics and
chemistry of the doping blocks \cite{recipe}. In this short note we
adress the specific role of the metal-oxygen complexes present in the
doping blocks of the superconducting Cu, Tl, Hg, Bi, Pb copper-oxides.
\section{Atomic sites of the dopants}
The crystal structures of the superconducting oxides are characterized
by superconducting $\rm CuO_2$ sheets embedded in a dielectric medium,
which consists of structurally and electronically different block
layers. The $\rm CuO_2$ layers are structurally nearly perfect, whereas
the block layers exhibit many structural imperfections, which are also a
source of electronic inhomogenities. The dopants reside in atomic sites
adjacent to the $\rm CuO_2$ planes \cite{Mue93}. However, it is
difficult to localize these sites precisely. Anions (mostly oxygen) and
and a few species of cations act as dopants, independently of each
other, or in combination, {\it e.g.} in complexes with variable
oxidation states of the metals. Table \ref{tablestructure} lists a
compilation of superconducting copper-oxides, using a recently proposed
four number naming scheme \cite{ShaJor}. This scheme allows for an
immediate identification and comparisons of the different structures and
sub-structures. It is advantegous over the popular naming scheme
delineating the cation stoichiometries and rather helpful in localizing
the dopants and their sites. For example the ''123'' system writes in
the four number $(ijkl)$ naming scheme 1212 , where the first entry
labels $i=1$ 'insulating' layer (copper-oxygen chains), the second entry
$j=2$ 'spacing' layers (barium-apical oxygen), the third entry $k=1$
'separating' layer (rare earth), and the fourth entry $k=2$ 'conducting'
$\rm CuO_2$ layers \cite{Unified}.

In the sub-group lacking an 'insulating' layer (first entry: $i= 0$) the
dopants are located in or close to the two 'spacing' layers with
commonly barium, strontium and lanthanum as metal atoms. For example in
$\rm La_{2-x}Ba_xCuO_4$ the cation $\rm Ba^{2+}$ acts as dopant
replacing $\rm La^{3+}$ in the 'spacing' layer. In superoxygented $\rm
La_2CuO_{4+\delta}$ extra oxygens between the two adjacent $\rm
La^{3+}$-O 'spacing' layers dope the system. The so-called infinite
layer compounds are reported to exhibit in the $i=0$ sub-group the
highest $T_c\sim 90-130$ K. From the structural phenomenology they
appear as rather exceptional cases with $i.$ two divalent metals in the
'spacing' layer, $ii.$ no apical oxygen in the 'spacing' layer, and
$iii.$ an ill defined oxygen stoichiometry. Indeed, recent structural
investigations analyze the infinite-layer systems as phase mixtures, and
find the superconductivity in the $\rm (Sr,Ca)CuO_2$ system to come from
the tetragonal phases $\rm Sr_{n+1}Cu_{n}O_{2n+1+\delta}$ ($n=2,3,4$),
not from the infinite-layer phase \cite{ShaPay}. The 'spacing' layers of
$\rm Sr_{n+1}Cu_{n}O_{2n+1+\delta}$ ($n=2,3,4$) comprise the apical
oxygens but surprisingly no cations, since the strontium site is
assigned to the 'separating' layers. The lack of single phase compounds
$\rm Sr_{n+1}Cu_{n}O_{2n+1+\delta}$ prevents so far more detailed
characterizations of the dopant and its site. $\rm
Sr_{n+1}Cu_{n}O_{2n+1+\delta}$ is an interesting system, which possibly
exhibits a doping mechanism in the 'spacing' layer by excess oxygens
similar to superoxygenated $\rm La_2CuO_{4+\delta}$, but with enhanced
$T_c$.

The solution of the infinite-layer enigma \cite{ShaPay} supports
strongly the role of the apical oxygen as a necessary and a crucial
atomic constituent of the superconducting copper-oxides \cite{BisTes}.

Superconductivity in the $i=0$ system $\rm
Sr_{n+1}\-Cu_{n}\-O_{2n+1+\delta}$ also evi\-den\-ces that the
'insu\-lating layers' are not the decisive atomic constituents for
super\-conducting cop\-per-oxides with enhanced $T_c$. But 'insulating
layers' are able to stabilize single phases with enhanced $T_c$, which
is a problem interesting in itself.
\section{The cations in the 'insulating' layer}
Only a few species of metal atoms are found in the 'insulating' layers:
$\rm Cu^{3+,1+}, Tl^{3+}, Hg^{2+}, Bi^{3+}, Pb^{2+}$. In inorganic
chemistry these metal ions are classified as 'soft', in other words as
highly polarizable. High polarizability may be related to various
properties as large ionic radii, electronegativity, ionization
potentials.

But high polarizability has less to do with these properties than with
the electronic configurations in the actual ionization state. Ions
with a high density of low-lying electronic states exhibit
a strong tendency to enhanced polarizabilities, due to hybridization.
Hybridization as a source of high polarizabilites is well known in
ferroelectric oxidic perovskites \cite{MigBau} where $\alpha_{oxygen}$
is enhanced by hybridization of the transition metal $d$\/ states with
the oxygen $p$\/ states. The metal ions in the 'insulating' layers have
in common either closed shell $d^{10}$, or inert pair $s^2$\/
configurations.

$\rm Tl^{3+}, Hg^{2+}$ exhibit closed $d^{10}$ shells as well as $\rm
Cu^{1+}$. The electronic configuration of $\rm Cu^{3+}$ is usually
believed to be $3d^8$, but the $2p$ core-level spectra of excited
'trivalent' $d$\/ states exhibit the typical signatures of $\rm Cu^{1+}$
and $\rm Cu^{2+}$ ions. This finding is not surprising from the typical
monovalent- and divalent-like bondlengths in the distorted 'trivalent'
$\rm CuO_3$ plaquettes. We therefore assume $\rm Cu^{3+}$ has to have at
least partially closed shell $d^{10}$\/ properties.

$\rm Bi^{3+}$ and $\rm Pb^{2+}$ exhibit so-called inert pair $s^2$
configurations \cite{Kre}.

Both configurations, closed shell $d^{10}$\/ and inert pair $s^2$\/,
exhibit a strong tendency to polarize the oxygen environment. Their
stereochemistry is rather unusual inducing in the metal-oxygen complexes
anisotropic polarizibilities. For the interesting anorganic chemistry
of these metallic ions see $e.g.$ Ref. \cite{Kre} and references therein.
\subsection{Closed shell $d^{10}$ ions}
Closed shell $d^{10}$ ions are known to form compounds with a low
coordination number, and in particular linear complexes
\cite{Jor57,Org58}. The capability of $d^{10}$ ions to form linear
complexes is strongly correlated with excitations from the
$(n-1)d^{10}$\/ ground state to low lying $(n-1)d^9ns$ and
$(n-1)d^9np$\/ states.

$d_{z^2}$ and $s$\/ orbitals can be hybridized to give new orbitals of
the type $\alpha\psi(d_{z^2})+\beta\psi(s)$. If the two electrons, which
occupy the $d_{z^2}$ orbital of the free ion are put into the hybridized
$1/\sqrt{2}(d_{z^2}+s)$ orbital, the atom develops a strong
electronegativity in the $x$\/- and $y$\/-directions, while if the
hybridized $1/\sqrt{2}(d_{z^2}-s)$ orbital is occupied, a strong
electronegativity in $z$\/-direction is developed. Hence two
arrangements can occur: {\it i.} two electrons in the
$1/\sqrt{2}(d_{z^2}-s)$ orbital and two short bonds in $z$\/-direction,
and {\it ii.} two electrons in the $1/\sqrt{2}(d_{z^2}+s)$ orbital with
four bonds in the $x$\/- and $y$\/- directions. Most of the closed shell
$d^{10}$ insulating Cu compounds, {\it e.g.} $\rm Cu_2O, LaCuO_2, MCuO
(M=Na, Rb, K)$, exhibit the arrangement {\it i.} with two short Cu--O
bonds of about 1.84 $\rm \AA$ in $z(c)$\/-direction. The same occurs in
the 'insulating' layer of $\rm RBa_2Cu_3O_{7-\delta}$ with two short
bonds (1.85 $\rm \AA$) along the $c$\/-axis to the apical oxygen, and
longer bonds (1.95 $\rm \AA$) in the $y(b)$\/ direction (chains). Also
the short bonds of the Tl-, Hg-complexes point in the $z(c)$\/-direction
\cite{BouFle}, whereas the long bonds form diagonals in the
$x(a)-y(b)$\/ plane.

Due to the hybridized $1/\sqrt{2}(d_{z^2}-s)$ orbital $d^{10}$ ions
destabilize octahedral and stabilize linear evironments. Preferentially
the $d^{10}$ ions with the lowest $nd^9(n+1)s$ excitation energy form
 linear complexes, because the energy required for $d-s$ mixing is
proportional to the $d-s$ separation \cite{Jor57}. But large $d-s$
separations are expected to favour tetrahedral or octahedral
environments through $s-p$ hybridization. Table \ref{tableenergy}
compiles the energies of lowest $d^9s$ and $d^9p$ states above the
$d^{10}$ state of closed shell $d^{10}$ ions. $\rm Zn^{2+}$ and $\rm
Cd^{2+}$ exhibit the largest $d-s$ separations, which may explain their
tendency to form tetrahedral or octahedral environments in
nonsuperconducting oxides. Well known Zn inserted in $\rm
RBa_2Cu_3O_{7-\delta}$ does not adapt to the linear complex in
the 'insulating' layer, substitutes preferentially for Cu in the
'conducting' layer, and suppresses efficiently $T_c$. Also Cd-oxygen
complexes were found to deteriorate superconductivity in copper-oxides.
$\rm Cu^{1+,3+}, Hg^{2+}, Tl^{3+}$ have smaller $d-s$ separations, form
 linear complexes and stabilize superconductivity. From the small $d-s$
separations $\rm Au^{1+}$- and $\rm Ag^{1+}$-ions might be be regarded
as a particularily well suited metallic species for the formation of
linear oxygen complexes. Au was found to substitute for Cu in the chains
of $\rm YBa_2Cu_{3-y}Au_yO_{7-\delta}$ up $y\sim 0.1$, enhancing
slightly $T_c$ \cite{EibFel}. At higher concentration Au-ions tend to
form metallic clusters due to $d^{10}$-$d^{10}$ interactions \cite{Kre}.
Silver-oxides decompose already at moderate temperatures far below the
formation of the copper-oxide. Hence high-pressure synthesis might be be
a possible route for synthesis \cite{Jan}.
\subsection{Inert pair $s^2$\/ ions}
Chemical bonding in complexes with inert pair $s^2$\/ ions is largely
determined by $s-p$ mixing \cite{Kre,Org58}. In $\rm Bi^{3+}$ (and $\rm
Pb^{2+}$) $6s-6p$\/ hybridization results in a pair of electrons
(lone-pair or inert pair) being pushed off to one side of Bi so that the
strong bonds are on the other side. Thus the unshared $s^2$\/ electron
pair causes a tendency to instability to antisymmetric distortions in
the cubically co-ordinated ion. The corresponding anisotropic
polarizibility tends to remove the center of inversion from the system.
Even if the cubic complex is stable it is particularily susceptible to
electrical polarization. $s-p$ hybridization of the inert pair $s^2$\/
ion $\rm Bi^{3+}$ may be therefore responsible for the large dielectric
properties of the 'insulating' layer in the superconducting Bi
copper-oxides.

Pb form the most complicated metal-oxygen complexes with sometimes more
than one metal sites. For instance the 'insulating' layer in {\it e.g.}
$\rm Pb_2Sr_2YCu_3O_8$ comprises $\rm Pb^{2+}$ and $\rm Cu^{1+}$, see
Ref. \cite{ShaJor} and references therein. The
discussion of these metal-oxygen complexes is beyond the scope of this
paper.
\section{Conclusion}
'Insulating' layers and their metal-oxygen complexes stabilize single
phase copper-oxide superconductors with enhanced $T_c$. The stabilizing
mechanism may be related to the specific electronic structure of the few
'successful' metals, which in their actual oxidation states turn out to
be either closed shell $d^{10}$, or inert pair $s^2$ ions. Both species
have a strong tendency to polarize the oxygen environment due to $d-s$
or $s-p$ hybridization. In particular $d^{10}$ ions form linearily
coordinated oxygen environments strongly polarizing the apical oxygens.

Our study on the possible specific role of the cations Cu, Tl, Hg, Bi,
and Pb in the doping block strengthens statements \cite{BisTes} on the
very importance of strongly polarized apical oxygens for
superconductivity in copper oxides. $d_{z^2}-s$ hybridization implies
essentially an ionic bond in which $\rm Cu^{1+}$ forms a $d^8$ plus
doubly occupied $(3d_{z^2}-4s)^2$ hybrid state instead of the
spherically symmetric $d^{10}$ state. $d_{z^2}-s$ hybridization has been
experimentally \cite{ResSch} and theoretically \cite{MarSch} confirmed
to give rise to the linear Cu-O complexes in semiconducting $\rm Cu_2O$
(cuprite). Hence $\rm Cu_2O$ may be considered as the archetypal closed
$d^{10}$ shell copper-oxide allowing for 'clean' low temperature studies
of the copper-oxide $3d_{z^2}-4s$ hybrid. Interestingly the electrical
conductivity of $\rm Cu_2O$ exhibits so far unexplored anomalies at
about 280, 200, 160, 130, 110 and 90 K \cite{BlaSch}.
\end{multicols}
\newpage
%
%
\begin{table}
\caption{Structure types of some typical superconducting copper-oxides.
The first column lists the generic assignment in the four number naming
scheme (see text). The second column indicates common assignments.
Column $i$ lists the possible doping ions in the 'insulating' layer.
Column $j$ lists the possible doping ions in the 'spacing' layer.}
\label{tablestructure}
\begin{tabular}{lllrll}
$ijkl$&Common&Chemical Formula\tablenotemark[1]&   $T_c$ [K]&$i$&$j$\\
\tableline
0201&T&$\rm (LaBa)_2CuO_4$&38&&$\rm La^{3+}, Ba^{2+},O^{2-}$\\
0201&T&$\rm La_2CuO_{4+\delta}$&44&&$\rm O^{2-}$\\
0201&T'&$\rm (Nd,Ce)_2CuO_{4-\delta}$&24&&$\rm Nd^{3+},Ce^{3+,4+},
O^{2-}$\\
0201&$\rm T^*$&$\rm (Nd,Sr,Ce)_2CuO_4$&35&&$\rm
Nd^{3+},Ce^{3+,4+},Sr^{2+},O^{2-}$\\
0212&&$\rm (La,Sr,Ca)_3Cu_2O_6$&58&&$\rm La^{3+},Sr^{2+},Ca^{2+},
O^{2-}$\\
0223&&$\rm (PbBa)(YSr)Cu_3O_8$&50&&$\rm Pb^{?},Ba^{2+},O^{2-}$\\
0234&&$\rm (Sr,Ca)_5Cu_4O_{10}$&70&&$\rm Sr^{2+},Ca^{2+},O^{2-}$\\
02''$\infty$\/-1''$\infty$&&$\rm (Ca,Sr)CuO_2$
&-\tablenotemark[2]&&$\rm Ca^{2+},Sr^{2+},O^{2-}$\\
0223&&$\rm Sr_4Cu_3O_{7+\delta}$&90-110&&$\rm O^{2-}$\\
\tableline
1201&&$\rm HgBa_2CuO_{4+\delta}$&98&$\rm Hg^{2+},O^{2-}$&\\
1212&123&$\rm YBa_2Cu_3O_{7-\delta}$&92&$\rm Cu^{3+,1+},O^{2-}$&\\
1223&&$\rm TlBa_2Ca_2Cu_3O_{9+\delta}$&123&$\rm Tl^{3+},O^{2-}$&\\
\tableline
2201&&$\rm Tl_2Ba_2CuO_6$&95&$\rm Tl^{3+},O^{2-}$&\\
2212&124&$\rm YBa_2Cu_4O_8$&80&$\rm Cu^{1+,3+},O^{2-}$&\\
2234&&$\rm Tl_2Ba_2Ca_3Cu_4O_{12}$&112&$\rm Tl^{3+},O^{2-}$&\\
2223&&$\rm Hg_2Ba_2Ca_2Cu_3O_{10}$&133&$\rm Hg^{2+},O^{2-}$&\\
\tableline
3201&&$\rm Pb_2(Sr,La)_2Cu_2O_6$&32&$\rm Pb^{2+},Cu^{1+}$&$\rm
Sr^{2+},La^{3+},O^{2-}$\\
3212&&$\rm Pb_2Sr_2YCu_3O_8$&70&$\rm Pb^{2+},Cu^{1+}$&\\
\end{tabular}
\tablenotemark[1]{For references of the structural data see $e.g.$
Ref. \cite{ShaJor}.}
\tablenotemark[2]{see text.}
\end{table}
%
%
%
\begin{table}
\caption{Energies of lowest $d^9s$ and $d^9p$ states
above the $d^{10}$ states (in $\rm cm^{-1}$)}
\label{tableenergy}
\begin{tabular}{lccccccc}
&$\rm Cu^{1+}$&$\rm Zn^{2+}$&$\rm Ag^{1+}$&$\rm Cd^{2+}$&$\rm Au^{1+}$
&$\rm Hg^{2+}$&$\rm Tl^{3+}$\\
\tableline
$nd^9(n+1)s$&21928&76105&39164&80463&15039&42862&75052\\
$nd^9(n+1)p$&66418&137876&80173&139042&63052&118616&147635\\
\end{tabular}
\tablenotemark{from Ref. \cite{Org58}}
\end{table}

\begin{references}
\bibitem{Unified}
The unified picture of superconductivity in oxides suggested in Ref.
\cite{BisTes} labels the rare earth or Ca layers between the metallic
$\rm CuO_2$ layers as 'insulating'. Here we follow for convenience the
four number naming scheme labeling the metal-oxygen complexes in the $P$
layer of Ref. \cite{BisTes} as 'insulating', and the 'insulating' layers
of Ref. \cite{BisTes} as 'separating'.
\end{references}
\end{document}